\documentclass[12pt]{iopart}

\usepackage{iopams}

\usepackage{graphicx} 
\usepackage{dcolumn} 
\usepackage{bm} 
\usepackage{textcomp}

\usepackage{color}
\usepackage{braket}

\begin{document}
\title{Single-site-resolved imaging of ultracold atoms in a triangular optical lattice}

\author{
Ryuta Yamamoto$^1$,
Hideki Ozawa$^1$,
David C. Nak$^{1,2}$,
Ippei Nakamura$^1$
\footnote{Present address: Kanagawa Institute of Industrial Science and Technology (KISTEC), Ebina 243-0435, Japan.}
and Takeshi Fukuhara$^1$}
\address{$^1$RIKEN Center for Emergent Matter Science (CEMS), Wako 351-0198, Japan}
\address{$^2$Institut f\"{u}r Laserphysik, Universit\"{a}t Hamburg, 22761 Hamburg, Germany}
\ead{takeshi.fukuhara@riken.jp}

\begin{abstract}
We demonstrate single-site-resolved fluorescence imaging of ultracold $^{87}\mathrm{Rb}$ atoms in a triangular optical lattice by employing Raman sideband cooling.
Combining a Raman transition at the D1 line and a photon scattering through an optical pumping of the D2 line, we obtain images with low background noise.
The Bayesian optimisation of 11~experimental parameters for fluorescence imaging with Raman sideband cooling enables us to achieve single-atom detection with a high fidelity of $(96.3 \pm 1.3)$\%.
Single-atom and single-site resolved detection in a triangular optical lattice paves the way for the direct observation of spin correlations or entanglement in geometrically frustrated systems.
\end{abstract}

\noindent{\it Keywords}: cold atoms, optical lattice, laser cooling, quantum gas microscope, frustrated magnetism, machine learning

\maketitle

\section{Introduction}
Frustrated quantum magnets are one of the most challenging subjects in condensed matter physics.
The simplest example of frustrated magnetism is a triangular lattice with antiferromagnetic interactions.
A competition of the interactions among the spins often leads to exotic liquid states, in which the spins are strongly correlated while their long-range order is suppressed~\cite{LBalents:2010}. 

Quantum simulation with ultracold atoms in optical lattices~\cite{IBloch:2012, CGross:2017} is a promising approach for understanding such systems.
The simulation of classical magnets of frustrated systems has been demonstrated using Bose--Einstein condensates in a triangular optical lattice~\cite{JStruck:2011}.
Further, theoretical proposals for the quantum simulation of frustrated quantum magnetism have been reported~\cite{AEckardt:2010,YChen:2010,DYamamoto2019,DYamamoto:2020}.
However, several technical challenges, such as the preparation of a low-entropy state, need to be overcome to realise frustrated quantum magnets in ultracold-atom experiments. 

Recently, there have been considerable advances in the simulation of quantum magnetism using the quantum gas microscope technique~\cite{CGross:2017}.
An antiferromagnetic order was realised in the Fermi--Hubbard model on a square lattice~\cite{AMazurenko:2017}.
To achieve a low-entropy state, which is essential for observing a magnetic order or correlation, the manipulation of local potential has been demonstrated~\cite{CChiu:2018}. 
In addition to two-spin correlations, string order~\cite{MEndres:2011,THilker:2017} and entanglement entropy~\cite{RIslam:2015} are accessible; such measurements provide a unique probe of quantum spin liquid states~\cite{LSavary:2017}.
Exotic excitations such as spinons, arising from quantum spin liquid states, may be captured through direct observation of spin dynamics~\cite{TFukuhara:2013a, TFukuhara:2013b,TFukuhara:2015}.

In this study, we demonstrate the single-site-resolved imaging of ultracold $^{87}\mathrm{Rb}$ atoms in a triangular optical lattice.
Although polarisation gradient cooling has been successfully employed to realise a quantum gas microscope of $^{87}\mathrm{Rb}$ atoms for a square lattice~\cite{WBakr:2009,JSherson:2010}, we used Raman sideband cooling, which has been applied to microscopes of fermionic $^{6}\mathrm{Li}$ and $^{40}\mathrm{K}$ atoms~\cite{LCheuk:2015,MParsons:2015,AOmran:2015}.
The reasons for which are as follows.
Generally, it is challenging to provide optical access for polarisation gradient cooling beams, except in the square lattice case, where one can irradiate the cooling light by superimposing optical lattice beams~\cite{JSherson:2010}.
From the perspective of optical access, Raman sideband cooling is relatively easy to implement.
Furthermore, in our scheme, scattered light from the cooling beams near the D1 line can be removed using an optical filter, and thus the fluorescence photons of the D2 line are detected with low stray light.
A drawback of this scheme is the relatively large number of experimental parameters required for the cooling method.
However, we have tuned them efficiently using machine learning based on Bayesian optimisation~\cite{PWigley:2015,BHenson:2018,INakamura:2019,EDavletov:2020,ABarker:2020}.

\section{Experimental setup and atom preparation}
We first describe the apparatus used for preparing and detecting ultracold $^{87}\mathrm{Rb}$ atoms in a triangular optical lattice.
To set up a high spatial resolution imaging system compatible with a laser cooling system, we employed the optical transport technique~\cite{TGustavson:2002}.
The laser-cooled atoms were directly loaded into an optical dipole trap generated by a laser beam at a wavelength of 817~nm.
The power of the beam was 450~mW, and the beam waist was 30~\textmu m.
To transport the atoms to the imaging region above a high numerical aperture (NA) objective of 0.65 (Mitsutoyo M Plan Apo NIR HR50x (custom)), we shifted the focus position of the trap beam by 119~mm in 0.9~s with an air-bearing moving stage.
Subsequently, the atoms were loaded into a crossed dipole trap generated by two beams with a beam waist of $40$~\textmu m and a wavelength of 1064~nm.
Here the internal atomic state was randomly populated owing to photon scattering induced by the transport beam.
To efficiently cool atoms down by evaporative cooling, the atoms were initialised into the $\ket{F=2, m_F=-2}$ state via optical pumping using two circularly polarised beams.
Then, we conducted evaporative cooling by lowering the potential depth of the crossed dipole trap.
The typical temperature of the atoms after the evaporative cooling was approximately 1~\textmu K.

\begin{figure}[t!]
    \centering
    \includegraphics[width=0.8\linewidth]{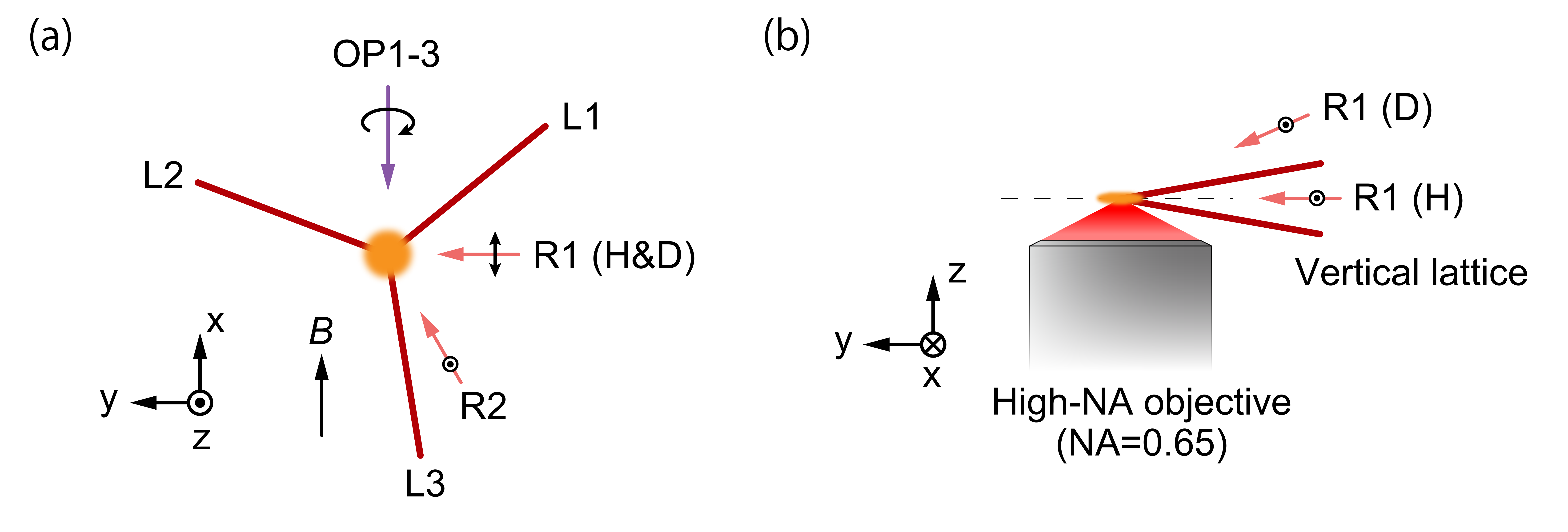}
    \caption{
    (a) Top view of the high-resolution imaging system.
    L1, L2, and L3 denote the triangular lattice beams; R1(H/D) and R2 denote the Raman beams; and OP1, OP2, and OP3 denote the optical pumping light with $\sigma^-$ polarisation.
    The angles between lattice beams L1--3 are $120^\circ$.
    The polarisation of R1(H/D) is parallel to the $x$-axis and corresponds to $\pi$ polarisation.
    The polarisation of R2 is oriented along the $z$-axis and consists of $\sigma^+$ and $\sigma^-$ polarisation.
    The applied magnetic field $B$ is 1~G along the $x$-direction.
    (b) Side view of the high-resolution imaging system.
    The angle between the vertical lattice beams is approximately $18^\circ$.
    R1(H) and R2 are propagating on the $xy$-plane.
    R1(D) is irradiated at an angle of $24^\circ$ from the $xy$-plane.
    Spontaneously scattered photons, resulting from the optical pumping beams (OP1--3), are collected through a high-NA objective of 0.65.
    }
    \label{fig:setup}
\end{figure}

We loaded the ultracold atoms into a lattice system that consisted of a triangular and a vertical optical lattices (see \fref{fig:setup}(a) and (b)).
We first ramped up the vertical optical lattice (along the $z$-direction), which was formed by two 810~nm beams at a relative angle of approximately $18^\circ$, as illustrated in \fref{fig:setup}(b).
The waist of the beams was approximately $(100, 43)$~\textmu m along the horizontal and vertical axes, respectively.
The lattice separation of the vertical lattice was $2.58$~\textmu m, which is smaller than the atomic cloud size in the optical trap.
Thus, the atoms were distributed over several layers of the vertical lattice.
Under this circumstance, we could not obtain clear images because of the contribution of considerably blurred fluorescence signals from the out-of-focus layers.
To obtain a site-resolved image, we prepared atoms trapped in a single layer using the following procedure.
We applied a magnetic field gradient of 21.4~G/cm along the $z$-direction.
The position-dependent Zeeman shift caused by this gradient made it possible to resolve each layer of the vertical lattice with a microwave transition.
The atoms trapped in the selected layer were transferred to the $\ket{F=1, m_F=-1}$ state, and then the other atoms in the $\ket{F=2, m_F=-2}$ state were blown away by a laser pulse resonant to the transition from $F=2$ to $F'=3$.
Finally, we ramped up the triangular optical lattice formed by three 1064~nm beams, intersecting at a relative angle of $120^\circ$ on the horizontal $xy$-plane (L1--3), as depicted in \fref{fig:setup}(a).
Notably, the shape of the triangular lattice beams was elliptical for reducing the harmonic confinement, and the typical beam waist was $(130, 44)$~\textmu m along the horizontal and vertical axes, respectively.
For site-resolved imaging in the triangular lattice, we rapidly increased the potential depths of the triangular and vertical lattices to (120, 170)~\textmu K, respectively, and pinned the atoms to their lattice sites.

\section{Fluorescence imaging with Raman sideband cooling}
\begin{figure}[t!]
    \centering
    \includegraphics[width=0.8\linewidth]{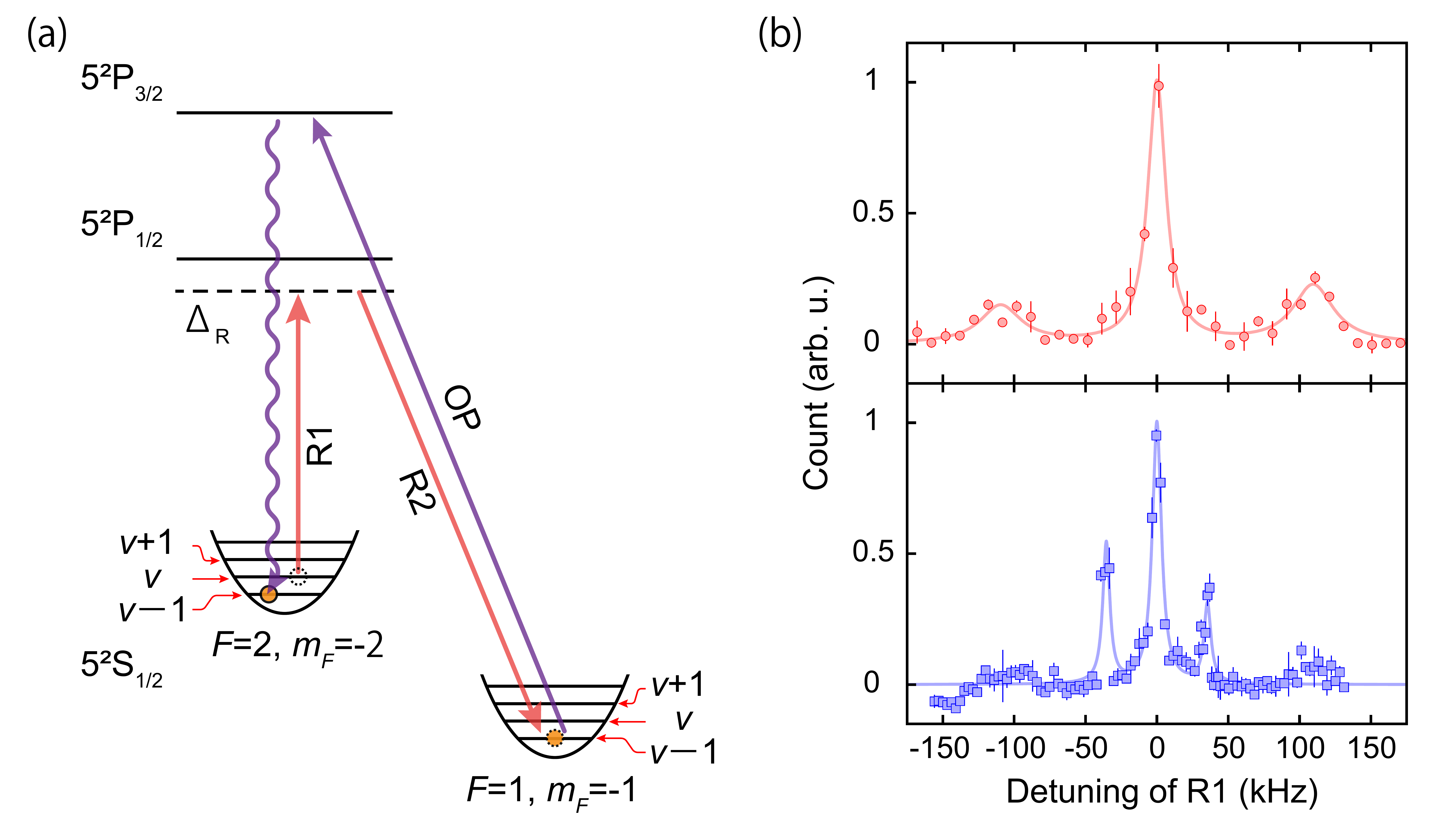}
    \caption{
    (a) Raman sideband cooling scheme.
    Raman beams R1(H/D) and R2 near the D1 line drive a transition from the $\ket{F=2, m_F=-2, \nu}$ state to the $\ket{F=1, m_F=-1, \nu-1}$ state.
    The atoms in the $\ket{F=1, m_F=-1, \nu-1}$ state are optically pumped back to the $\ket{F=2, m_F=-2, \nu-1}$ state through optical pumping~(OP).
    The detuning $\Delta_R$ of the Raman transition from the intermediate state is $-100$~GHz.
    (b) Raman sideband spectra.
    The top and bottom panels present typical Raman spectra obtained by applying R1(H)-R2 and R1(D)-R2 pairs, respectively.
    The lines are fits to Lorentzians for the carrier and sidebands. 
    The differences between the carrier and sidebands are $(108, 35)$~kHz in the horizontal and vertical directions, respectively.
    The amplitude and offset of each spectrum are adjusted.
    }
    \label{fig:RSC}
\end{figure}

We adopted fluorescence imaging to achieve single-atom sensitivity.
However, light scattering during imaging causes the atoms to heat up, which leads to atom loss.
To suppress the undesired heating, we utilised Raman sideband cooling, which we describe here.
Raman beams R1 and R2, depicted in \fref{fig:RSC}(a), drive a transition that decreases vibrational levels from the $\ket{F=2, m_F=-2, \nu}$ state to the $\ket{F=1, m_F=-1, \nu-1}$ state.
Here $\nu$ is the vibrational level.
The atoms transferred to the $\ket{F=1, m_F=-1, \nu-1}$ state are optically pumped back to the $\ket{F=2, m_F=-2, \nu-1}$ state using optical pumping~(OP) light with $\sigma^-$ polarisation (see \fref{fig:RSC}(a)).
The OP1 (OP2) light transfers atoms from $F=1$ ($F=2$) to $F'=2$.
In the optical pumping process, the changes in the vibrational levels are suppressed by tightly trapping the atoms in potential wells up to the point of the Lamb--Dicke regime~\cite{DLeibfried:2003}.
Owing to the repetition of the above processes, the atoms are cooled down to the vibrational ground state $\ket{F=2, m_F=-2, \nu=0}$, whereas photons resulting from the optical pumping are used for fluorescence imaging.
The photons collected with the high-NA objective were recorded at an electron-multiplying CCD camera.
The atoms stop scattering photons when they populate the vibrational ground state.
For continuous fluorescence imaging, we applied an additional beam~(OP3), which transferred the atoms from $F=2$ to $F'=3$.

The wavelength of the fluorescence light induced by the optical pumping was 780~nm of the D2 line, whereas the Raman beams were detuned by $-100$~GHz from the D1 line at 795~nm. 
Therefore, the stray light from the Raman beams can be removed from the fluorescence light using an interference filter.
Furthermore, the optical pumping beams OP1--3 propagated perpendicularly to the objective axis, which reduced their stray light incident onto the imaging system (see \fref{fig:setup}).
Owing to these advantages, we obtained fluorescence images with only a few background photons.

We used two Raman beam pairs to independently adjust the coupling strengths of the Raman transition along the horizontal and vertical directions.
The R1(H) and R2 beams drove the Raman transition along only the horizontal direction, and the R1(D) and R2 beams drove the transition along mainly the vertical direction.
Here R1(H) propagated along the $y$-axis, and R1(D) was irradiated at an angle of $24^\circ$ from the horizontal plane (see \fref{fig:setup}).
To experimentally confirm the role of each Raman beam pair (R1(H)-R2 and R1(D)-R2), we took Raman spectra.
The spectra of the former and latter pairs are presented in the top and bottom panels of \fref{fig:RSC}(b), respectively.
The results of the Raman spectra indicate that the separations of the vibrational levels of the lattice potential wells were $(\omega_H, \omega_V) = 2\pi\times(108, 35)$~kHz in the horizontal and vertical directions, respectively.
We evaluated the Lamb--Dicke parameter using these separations.
For the spontaneous emission by the optical pumping, the Lamb--Dicke parameters, $\eta_{\mathrm{OP}}=k_{\mathrm{OP}}a$, were $0.19$ and $0.33$ in the horizontal and vertical directions, respectively, where $\hbar k_{\mathrm{OP}}$ is the momentum of the optical pumping beams, and $a=\sqrt{\hbar/2m\omega}$ is the harmonic oscillator length.
The Lamb--Dicke parameter for the Raman beams is described as $\eta_R = \Delta k a$, where $\hbar \Delta k$ is the momentum transfer due to the Raman beams.
Here the Raman momentum transfers for the R1(H)-R2 and R1(D)-R2 pairs are expressed as $\Delta k_H$ and $\Delta k_D$, respectively.
The momentum transfer $\Delta k_H$ included only the horizontal component of $9.0$~\textmu m$^{-1}$, whereas $\Delta k_D$ contained a horizontal component of $8.6$~\textmu m$^{-1}$ as well as a vertical component of $3.2$~\textmu m$^{-1}$.
Therefore, the Lamb--Dicke parameter, $\eta_H=\Delta k_H a$, is $0.21$ for the beam pair R1(H)-R2.
Similarly, for the beam pair R1(D)-R2, the Lamb--Dicke parameters, $\eta_D=\Delta k_D a$, are $0.20$ in the horizontal direction and $0.13$ in the vertical direction.
All the Lamb--Dicke parameters ($\eta_\mathrm{OP}$, $\eta_H$, and $\eta_D$) are significantly lower than 1, indicating that the atoms were in the Lamb--Dicke regime for the cooling process.

In the preceding discussion, we assume that each site of the triangular lattice is well approximated by a two-dimensional isotropic harmonic trap.
Under this assumption, there is no angular dependence in the coupling strength of the Raman transition on the horizontal plane.
However, considering the spatial symmetry of a triangular lattice, a relative angle of $15^\circ$ between the incident direction of a lattice beam and Raman momentum can provide the best Raman transfer efficiency for the triangular lattice.
Owing to the limitation of our apparatus, the relative angle was approximately $19^\circ$.
Although this angle was slightly different from the best one, we successfully cooled the atoms in the triangular lattice through Raman sideband cooling.

\section{Automatic optimisation of cooling parameters}
\begin{figure}[t!]
    \centering
    \includegraphics[width=1\linewidth]{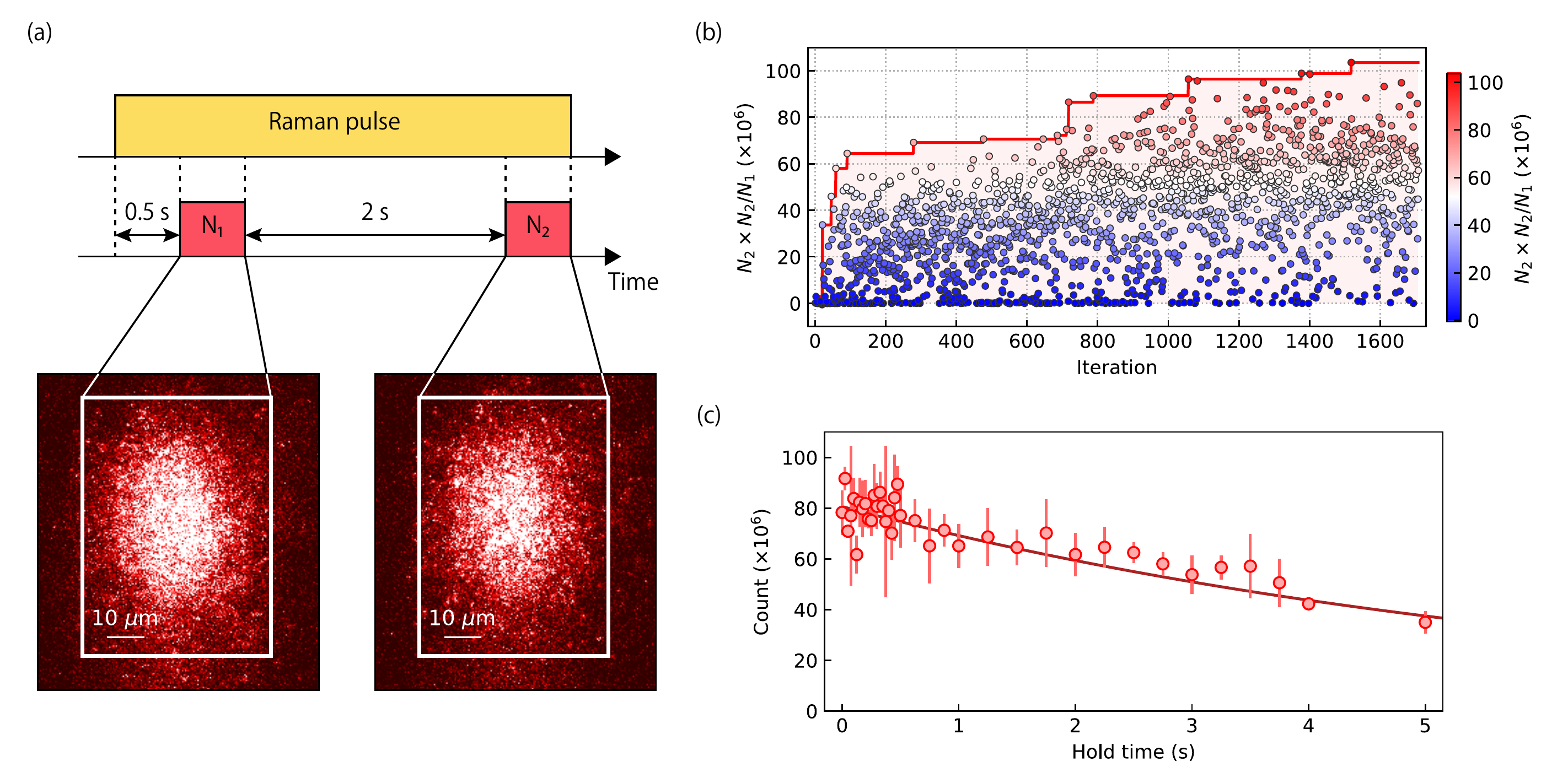}
    \caption{
    Parameter optimisation for Raman sideband cooling.
    (a) Sequence to evaluate the score of parameters selected by the Bayesian optimiser.
    A Raman pulse of 0.5~s is applied before taking the first image to remove the effect of initial loss induced by light-assisted collision.
    The exposure time of the two images is 0.5~s, and the wait time between the images is 2~s.
    The fluorescence count within the region indicated by the white box denotes $N_1$ and $N_2$ for the first and second images, respectively.
    (b) Temporal growth of optimisation score $N_2 \times (N_2/N_1)$.
    The red line represents the maximum score among all the searched results up to that trial.
    (c) Lifetime of fluorescence imaging.
    The fluorescence count with the 0.25~s exposure is plotted as a function of hold time.
    The lifetime with the optimised parameters is $(6.5 \pm 0.4)$~s.
    }
    \label{fig:MLO}
\end{figure}

To improve the quality of the fluorescence images, the laser parameters of the Raman beams and optical pumping beams need to be fine-tuned.
Because of experimental imperfections, such as the limited purity of the polarisation of the optical pumping beams or heating from the optical lattices, it is generally difficult to determine the desirable parameters without experimental optimisations.
Moreover, correlations between multiple parameters often compel experimenters to expend considerable time and effort.
We employed Bayesian optimisation to overcome this difficulty.
The details are explained in our previous work on Bayesian optimisation of evaporative cooling~\cite{INakamura:2019}.
To improve the signal-to-noise ratio of the fluorescence image, it is necessary to increase photon scattering, which inevitably leads to the heating of atoms.
To ensure high-fidelity detection, it is necessary to suppress hopping of the atoms, which also results in atom loss.
Therefore, Raman sideband cooling needs to sufficiently surpass the heating.
To this end, we adopted $N_2 \times (N_2/N_1)$ as the score to be maximised by the optimiser.
Here $N_1$ is the fluorescence count of the first image with an exposure time of 0.5~s, and $N_2$ is that of the second image with the same exposure time taken 2~s after the first image (see \fref{fig:MLO}(a)).
The factor $N_2$ of the score motivates the optimiser to increase the number of fluorescence photons.
The ratio $N_2/N_1$ corresponds to the lifetime of the atoms in the Raman sideband cooling process; less hopping is preferred to obtain a longer lifetime.
Under these conditions, the balance between increasing photon scattering and efficient cooling is maintained. 
Notably, our score based on the fluorescence counts is robust and easy to evaluate.
Furthermore, instead of using only the atoms in a single layer, we used those in all the layers to stabilise the atom numbers and increase the fluorescence count.

We optimised 11~parameters involving the intensities and frequencies of Raman beams and optical pumping beams.
The optimisation was conducted automatically and almost converged after 1700~trials within 13~h (see \fref{fig:MLO}(b)).
Therefore, this method can also be used to compensate for the slow drift of the experimental environment, such as the temperature change over the year or deterioration of optical components with age, and even to recover from slight accidental changes in the experimental apparatus.

Based on the outcome of the Bayesian optimisation, the two-photon Rabi frequencies of the Raman transition on the carrier were set to $2\pi \times 1.1$~kHz for R1(H)-R2 and $2\pi \times 6.8$~kHz for R1(D)-R2.
Using the optimised parameters, we achieved a lifetime of $(6.5 \pm 0.4)$~s of atoms during fluorescence imaging (see \fref{fig:MLO}(c)).

\section{Site-resolved imaging}

\begin{figure}[ht!]
    \centering
    \includegraphics[width=0.8\linewidth]{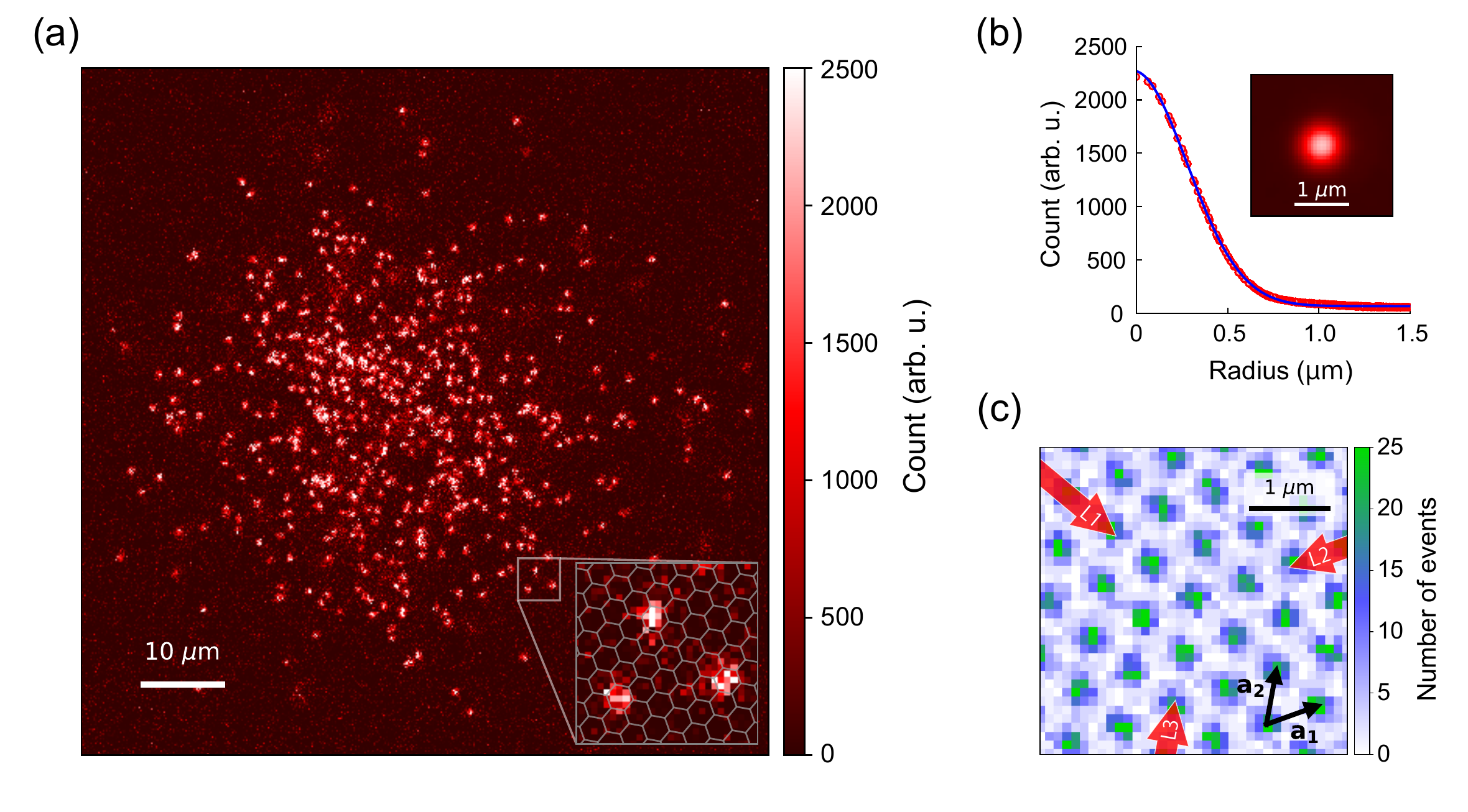}
    \caption{
    (a) Single-site-resolved fluorescence image of $^{87}\mathrm{Rb}$ atoms in a triangular optical lattice.
    The grey lines in the inset indicate the boundaries of each lattice site.
    (b) Radially averaged point spread function~(PSF) extracted from isolated atoms of fluorescence images.
    The blue curve represents the fit by a Gaussian. 
    The obtained full width at half maximum is 679~nm, which corresponds to an effective NA of 0.59.
    The inset presents the PSF averaged using subpixel shifting~\cite{WBakr:2009}.
    (c) Two-dimensional histogram of relative positions of atoms.
    The red arrows with labels L1--3 denote the incident direction of triangular lattice beams.
    The black arrows, $\bm{a}_1$ and $\bm{a}_2$, represent the lattice vectors of our lattice system.
    Using these lattice vectors, each lattice site can be expressed as $n \bm{a}_1 + m\bm{a}_2$, where $n$ and $m$ are integers.
    }
    \label{fig:psf_geom.}
\end{figure}

Using the optimised parameters explained in the preceding section, we realised single-site-resolved imaging of $^{87}\mathrm{Rb}$ atoms in a single-layer triangular optical lattice, as exhibited in \fref{fig:psf_geom.}(a).
The performance of a quantum gas microscope is characterised by the detection fidelity, which consists of pinning, hopping, and loss rates.
To evaluate these rates, the atom distribution needs to be determined, which requires the profile of the point spread function~(PSF) and lattice geometry, as characterised by the lattice vectors, $\bm{a}_1$ and $\bm{a}_2$.

First, we evaluated the PSF of our imaging system by averaging over 2500~isolated atoms of the fluorescence images (depicted in the inset of \fref{fig:psf_geom.}(b)).
The radial profile of the measured PSF is presented in \fref{fig:psf_geom.}(b).
We approximated the PSF as a Gaussian.
The fitting result is represented by the blue curve in \fref{fig:psf_geom.}(b); the full width at half maximum was determined as 679~nm, which corresponds to an effective NA of 0.59.
The obtained resolution of our imaging system was only 1.1~times larger than the diffraction-limited resolution of 617~nm, which is given by the designed NA of 0.65.
Converting the integrated PSF count into the number of photons, we found that $\sim 240$~photons per atom were detected during a 0.5~s exposure.
This count corresponds to a photon scattering rate of 7.8~kHz with an estimated collection efficiency of 6.0\%.

In the case of square lattices, the lattice vectors, comprising the lattice axis and spacing, can be determined by searching for a one-dimensional histogram with minimal width projected onto one axis~\cite{RYamamoto:2016}, because the lattice axes are orthogonal.
However, in our case of the triangular lattice, an additional coordinate transformation is necessary to extract the lattice vectors for the same method because the two lattice vectors are not orthogonal.
Therefore, we used a two-dimensional histogram of the relative coordinates between individual atoms (see \fref{fig:psf_geom.}(c)).
We fitted the two-dimensional histogram with the sum of Gaussians with the same amplitude $A$ and width $w$:
\begin{equation}
    H_{\mathrm{2D}}(\bm{r}) = A\sum_{n,m}\exp\left[-\frac{(\bm{r}-\bm{r}_{n,m})^2}{w^2}\right],
\end{equation}
where $\bm{r}_{n,m} = n\bm{a}_{1} + m\bm{a}_{2}$ indicates the centre coordinate of the lattice site $(n, m)$.
From the fitting result, we directly extracted information on the lattice vectors for our system.

\begin{figure}[t!]
    \centering
    \includegraphics[width=1\linewidth]{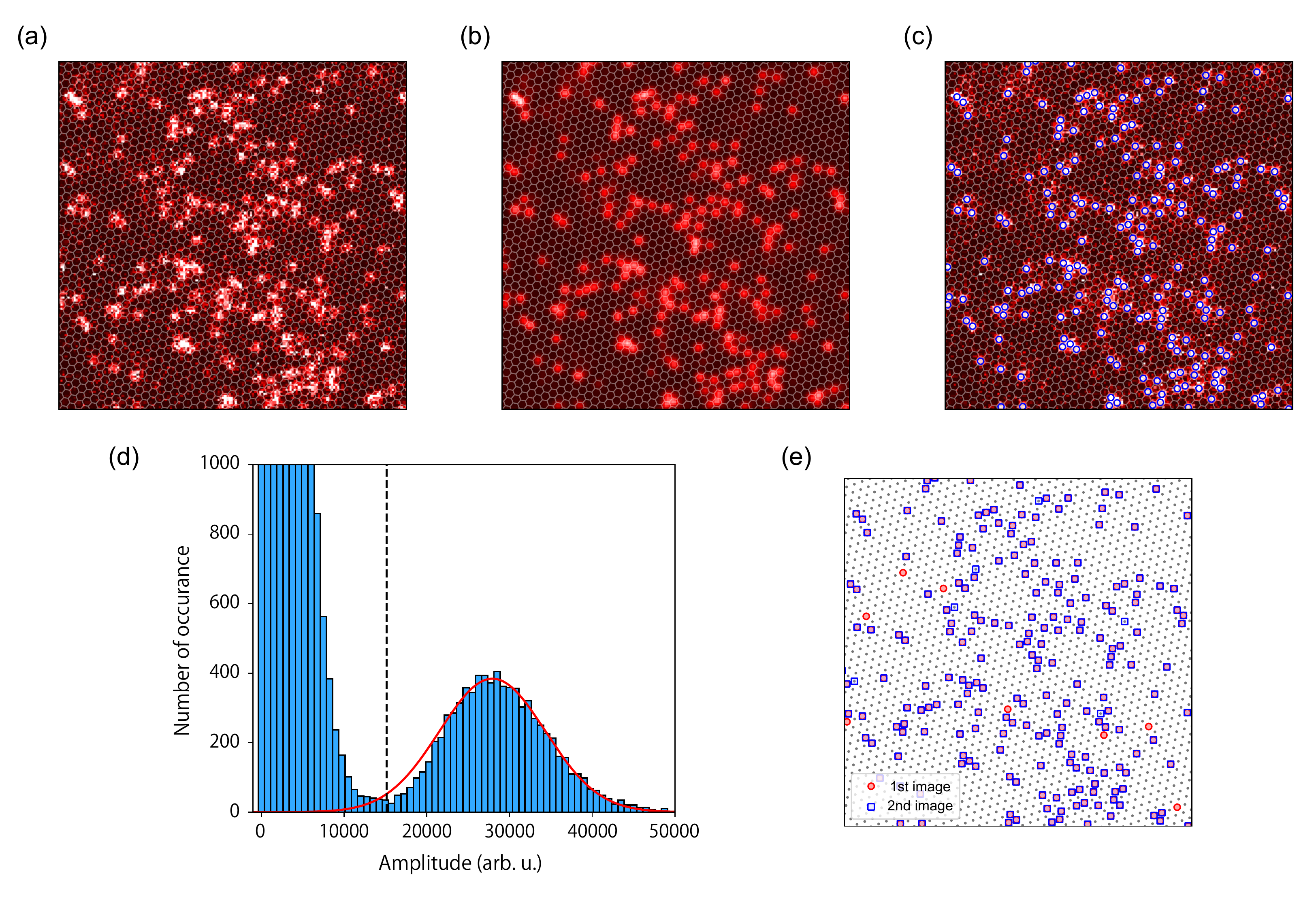}
    \caption{Reconstruction of atom distribution.
    (a) Image of the sparsely filled lattice within a limited region of $30 \times 30$~\textmu m.
    The exposure time is 0.5~s.
    (b) Reconstructed image.
    (c) Image with the blue circles indicating atom-occupied sites.
    The grey lines in (a)--(c) represent boundaries of each lattice site.
    (d) Histogram of reconstructed amplitude at the lattice sites.
    The red curve is a fit to the amplitude distribution of occupied sites through normal distribution with standard deviation $\sigma$.
    The threshold for the reconstruction is set to $2 \sigma$ lower than the centre of the normal distribution, represented by the dashed line.
    (e) Reconstructed atom distribution.
    The black dots indicate the lattice sites.
    The red circles and blue squares represent the atom-occupied sites of the first and second images, respectively.
    }
    \label{fig:fidelity}
\end{figure}

We determined the atom distribution in the images using a deconvolution method based on Richardson--Lucy deconvolution with the obtained PSF and lattice vectors $\bm{a}_1$ and $\bm{a}_2$~\cite{RYamamoto:2016}.
We applied this method to a limited region of $30\times30$~\textmu m, including approximately 2000~sites.
A typical example of the deconvolution process is presented in \fref{fig:fidelity}(a)--(c).
The histogram of the reconstructed amplitude at each site is shown in \fref{fig:fidelity}(d), in which the distribution of the empty and occupied sites are clearly separated.
We defined the threshold for the occupied site as $2 \sigma$ of the lower side in the distribution of occupied sites, which is indicated by the dashed line in \fref{fig:fidelity}(d).

Finally, we evaluated the detection fidelity by taking two images of the same atomic cloud and comparing their reconstructed atom distributions (\Fref{fig:fidelity}(e)).
In this measurement, the exposure time of each image was 0.5~s, and the wait time between the images was 0.15~s.
The pinning rate was defined as the rate of atoms that stayed pinned to their sites, the hopping rate as the rate of the occupied lattice sites that were detected only in the second image, and the loss rate as the ratio of the differences in atom number between the images.
Note that our definition of the loss rate allowed for an increase in the atom number due to atoms entering from outside of the analysis area; thus, the loss rate became negative sometimes.
We achieved a pinning rate of $(96.3 \pm 1.3)$\%, a loss rate of $(0.5 \pm 1.4)$\%, and a hopping rate of $(3.2 \pm 1.0)$\% for 0.5~s exposures of clouds with a lattice filling of approximately 0.10.

\section{Conclusion}
In conclusion, we have demonstrated the site-resolved imaging of ultracold $^{87}\mathrm{Rb}$ atoms in a triangular optical lattice.
Although we used the fluorescence counts of an atomic ensemble as the optimisation score, a high pinning rate of $(96.3 \pm 1.3)$\% during an imaging time of 0.5~s has been realised through the machine-learning-based automatic optimisation of the Raman sideband cooling parameters.
The detection fidelity can be improved further if the pining rate can be directly used as an optimisation score.
Thus, a more robust deconvolution algorithm is required.
Although the present experiment has been performed for $^{87}\mathrm{Rb}$ atoms, our method can be extended to $^{85}\mathrm{Rb}$ atoms, for which the interatomic interaction can be adjusted with a broad, low-field Feshbach resonance~\cite{SCornish:2000} for realising antiferromagnetic super-exchange coupling~\cite{AKuklov:2003,LDuan:2003} or antiferromagnetic coupling through negative absolute temperature~\cite{DYamamoto:2020}.

\ack
This work was supported by JSPS KAKENHI Grant Number JP19H01854 and ImPACT Program of Council for Science, Technology and Innovation (Cabinet Office, Government of Japan).
We acknowledge the stimulated discussion in the meeting of the Cooperative Research Project of the Research Institute of Electrical Communication, Tohoku University.

\section*{References}
\bibliographystyle{iopart-num}
\bibliography{references} 

\end{document}